\documentclass[hyper]{JHEP} 

\usepackage{epsfig}

\newcommand\fverb{\setbox\pippobox=\hbox\bgroup\verb}
\newcommand\fverbdo{\egroup\medskip\noindent%
			\fbox{\unhbox\pippobox}\ }
\newcommand\fverbit{\egroup\item[\fbox{\unhbox\pippobox}]}
\newbox\pippobox
\title{Proposal for the Open String  Tachyon Effective Action
 in the Linear Dilaton Background}
\author{by J. Kluso\v{n}\\
	 Department of Theoretical Physics and Astrophysics\\
                   Faculty of Science, Masaryk University\\
Kotl\'{a}\v{r}sk\'{a} 2, 611 37, Brno\\
Czech Republic\\
	E-mail: \email{klu@physics.muni.cz}}
\preprint{\hepth{0403124}} 
\abstract{In this paper 
we propose  tachyon effective actions
for unstable D-branes
in superstring and bosonic string theories 
in the presence of the linear dilaton
background.}

\def\bb{\mathbf{B}}

\def\ss{\sin \frac{\tau}{\sqrt{2}}}
\def\st{\sinh \frac{\tau}{\sqrt{2}}}
\def\st2{\sinh^2 \frac{\tau}{\sqrt{2}}}
\def\ss2{\sin^2 \frac{\tau}{\sqrt{2}}}

\begin{document}
\section{Introduction}\label{first}
It is well known that bosonic and
Type IIA and Type IIB string theories contain
in their spectrum unstable D-branes. 
These unstable D-branes are characterised
by having a single tachyonic mode
of the $\mathrm{mass}^2  \ \mu^2_{super}=-\frac{1}{2}$ in case
of supersymmetric non-BPS D-branes or
a tachyonic mode with $\mu^2_{bos}=-1$ 
(in $\alpha'=1$ unit) living on their
world-volume. The tachyon field has local
maximum at $T=0$ and pair of global minima at
$\pm\infty$ where the negative contribution
from the potential exactly cancels the tension
of non-BPS D-brane in supersymmetric theory, in
bosonic theory the story is basically the same
with the exception that there is only one global
minimum at $T=\infty$.  As a result the
configuration where the tachyon potential 
is at its minimum corresponds to vacuum without
any D-branes and fluctuations of the open
string field around these minima do not
contain any perturbative open string states
in the spectrum \footnote{For  review of the tachyon condensation
in string theory, see
\cite{Sen:1999mg,Taylor:2002uv,
Martinec:2002tz,Nakayama:2004vk,
Arefeva:2001ps,Ohmori:2001am,Schwarz:1999vu,
Lerda:1999um}. Some papers, where the effective
field theory descriptions of the tachyon
dynamics can be found, are
\cite{Sen:1999md,Garousi:2000tr,
Bergshoeff:2000dq,Kluson:2000iy,
Gibbons:2000hf,Kluson:2000gr,
Minahan:2000tg,Arutyunov:2000pe,Sen:2002nu,
Sen:2002an,Sen:2002qa,
Sen:2003tm,Kutasov:2003er,
Garousi:2003ai,Brax:2003rs,
Kim:2003ma,Sen:2003bc,Kluson:2003rd,
Kluson:2003sr,Smedback:2003ur,
Fotopoulos:2003yt,
Kluson:2003qk,Niarchos:2004rw}.}.

One can also study the time-dependent tachyon
solution where the tachyon rolls away from
near the top of the potential towards the minimum
of the potential at $T=\infty$
\cite{Sen:2002nu,Sen:2002an,Sen:2002qa,
Gutperle:2002ai,Sen:2002vv,Maloney:2003ck}. This time-dependent
tachyon condensation is described be marginal deformation
on the boundary of the world sheet $T(X^0)=e^{\mu  X^0}$.
It was shown in   \cite{Lambert:2003zr,Kutasov:2003er}
that this time-dependent tachyon condensation has  nice
effective field theory description based on
the superstring tachyon effective action 
\footnote{Our convention is $\eta_{\mu\nu}=
\mathrm{diag}(-1,1,\dots,1) \  , \mu \ , \nu=0,
\dots, D-1 \ , i , \ j=1,\dots, D-1$ where
$D$ is dimension of the space time.}
 \cite{Kutasov:2003er}
\begin{equation}\label{Kutact}
S=-\int d^Dx \mathcal{L} \ ,
\mathcal{L}=\frac{1}{g_s(1+\frac{T^2}{2})}
\sqrt{1+
\frac{T^2}{2}+\eta^{\mu\nu}
\partial_{\mu}T\partial_{\nu}T} \ ,
\end{equation}
where $g_s$ is constant string coupling constant
related to the constant dilaton $\Phi_0$ as 
$g_s=e^{\Phi_0}$. 
It was shown that this action correctly captures
the physics around the marginal tachyon condensation
in case of constant dilaton and the flat  
Minkowski space-time \cite{Kutasov:2003er,Niarchos:2004rw}. 
This action  was
reconstructed around the vicinity of the conformal
point corresponding to the time-dependent background 
which should represent the exact boundary conformal
field theory 
\begin{equation}\label{tachmarg}
T=T_+e^{\frac{x^0}{\sqrt{2}}}+T_-e^{-\frac{x^0}{\sqrt{2}}} \ .
\end{equation}
The remarkable fact about the action (\ref{Kutact}) is that
if we demand that the generic first order tachyon Lagrangian
\begin{equation}
\mathcal{L}=V(T)K((\partial T)^2)
\end{equation}
has (\ref{tachmarg}) as its exact solution fixes its 
time-dependent part to be 
\begin{equation}\label{Kutacttime}
\mathcal{L}=\frac{1}{g_s(1+\frac{T^2}{2})}
\sqrt{1+\frac{T^2}{2}-(\partial_0T)^2} \ .
\end{equation}
Then (\ref{Kutact}) arises from (\ref{Kutacttime}) 
by Lorenz-covariant generalisation which was extensively
discussed in \cite{Fotopoulos:2003yt}. Note
also that after field redefinition 
\begin{equation}\label{fieldred}
\frac{T}{\sqrt{2}}=\sinh\frac{\tilde{T}}{\sqrt{2}}
\end{equation}
the Lagrangian (\ref{Kutact}) becomes
"tachyon DBI" like Lagrangian
\begin{equation}\label{tachDBI}
\mathcal{L}=e^{-\Phi_0}V(\tilde{T})
\sqrt{-\det(\eta_{\mu\nu}+
\partial_{\mu}\tilde{T}
\partial_{\nu}\tilde{T}})=
e^{-\Phi_0}V(\tilde{T})\sqrt{1+
\eta^{\mu\nu}\partial_{\mu}\tilde{T}
\partial_{\nu}\tilde{T}} \ .
\end{equation}
Further support for the validity
of the (\ref{tachDBI}) in the description of the tachyon
dynamics around the marginal deformation was given in 
\cite{Sen:2003zf}.

Even if the relation between  the action (\ref{Kutact}) 
and the tachyon effective action calculated from the
string theory partition function is not completely clear
 \cite{Fotopoulos:2003yt}
we mean that the fact that the tachyon marginal
perturbation is an exact solution of the equation of motion
arising from
 (\ref{Kutact}) is very attractive and deserve further study. 
In particular, 
we can ask the question whether there are
another  marginal tachyon 
profiles that are    solutions of the equation
of motion arising from (\ref{Kutact}).
 Some interesting comments and suggestions
considering these problems
were also presented in \cite{Fotopoulos:2003yt} and we
hope to return to them in near future. 
For next purposes  we also present the version
of the D-brane tachyon effective action 
in bosonic string theory that 
we  proposed in 
\cite{Kluson:2003sr} 
\footnote{More detailed discussion of the
tachyon effective action in bosonic theory can
be found in \cite{Smedback:2003ur}.}
\begin{equation}\label{Kutbos}
\mathcal{L}=\frac{1}{g_s(1+T)}\sqrt{
1+T+\frac{\eta^{\mu\nu}\partial_{\mu}T
\partial_{\nu}T}{T}} \ .
\end{equation}
We have shown in \cite{Kluson:2003sr}
that the tachyon  profile
\begin{equation}
T=e^{\beta_{\mu}x^{\mu}} \ , 
 \beta_{\mu}\eta^{\mu\nu}\beta_{\nu}+1=0 \ 
\end{equation}
is  exact solution
of the equation of motion arising from
(\ref{Kutbos}).
The pure  time-dependent case
$\beta_0=1 \ , \beta_i=0$  corresponds to the decay of
unstable brane where in the far past D-brane is in
its unstable maximum $T=0$ and rolls to the stable
minimum $T=\infty$ at far future. 

Since the tachyon effective actions (\ref{Kutact}), 
(\ref{tachDBI}) and (\ref{Kutbos})
 are useful in the description of the 
tachyon dynamics near the marginal
tachyon profile in flat space time one can ask the
question whether their description of
the tachyon condensation in nontrivial closed background
is as successful as in flat space time. An example of 
such  a
closed string background that has simple
conformal field theory description and
where one can easily
find marginal boundary interaction 
\cite{Karczmarek:2003xm} 
 is linear dilaton background.
 The results given there were used recently 
in \cite{Kluson:2004pj} where 
we have shown that the tachyon
effective Lagrangian 
obtained from the string 
partition function is different
from the tachyon effective 
Lagrangian (\ref{Kutbos}). 
Despite this fact 
we mean that it is
 useful to try to find the 
generalisation of the action (\ref{Kutact})
to the linear dilaton background that will
not be directly related to the 
Lagrangian evaluated from the string
partition function but which will have 
the rolling tachyon profile in the linear
dilaton background as its exact solution.
 This is the question
which we will address in this paper.
We find such a form 
of the tachyon  action that
 reduces to (\ref{Kutact}) for constant
dilaton and  that  the rolling tachyon profile
in the linear dilaton background \cite{Karczmarek:2003xm}
is the exact solution of the equation of motion
that arises from it. 
We will also calculate the component of the stress energy tensor and
the dilaton source. Unfortunately we will find that 
their asymptotic behaviour
is different from the exact results 
given in \cite{Karczmarek:2003xm}. 
We will discuss this issue more extensively
in conclusion and suggest its possible
resolution.

The organisation of this paper is as follows. 
In  section
(\ref{second}) we present 
the main proposal for the tachyon effective
action in superstring theory. 
In section (\ref{third}) we 
will generalise the action (\ref{Kutbos}) 
to the case of linear dilaton
background as well. 
And finally in conclusion (\ref{fourth}) we 
outline our results and suggest possible 
extension of this work. 
\section{Proposal for supersymmetric
tachyon effective action in the linear
dilaton background}\label{second}
In this section we propose 
such a form of the tachyon
effective action in superstring theory
 which has the rolling tachyon
profile in the linear dilaton background as
its exact solution. The linear dilaton
background is characterised with 
the spacelike vector  $V_{\mu}$ so that
the dilaton is 
\begin{equation}
\Phi=V_{\mu}x^{\mu}+\Phi_0 \ ,
\end{equation}
where $\Phi_0$ is constant part of the dilaton and which
coincides with the constant string coupling constant $g_s$
in case of vanishing gradient $V_{\mu}=0$. 
Since it is not completely clear how to include
the massless scalars that parametrise transverse
position of unstable D-brane into the action 
(\ref{Kutact}) we restrict ourselves to the case
of D-brane that fills the whole space-time. 

The  motivation for  our proposal is
that the straightforward
generalisation of (\ref{Kutact}) and (\ref{Kutbos}) 
to the case of
the arbitrary dilaton $\Phi$
\begin{equation}\label{Kutdil} 
\mathcal{L}_{super}=\frac{e^{-\Phi}}{1+\frac{T^2}{2}}
\sqrt{1+
\frac{T^2}{2}+\eta^{\mu\nu}
\partial_{\mu}T\partial_{\nu}T} \  ,
\mathcal{L}_{bos}=\frac{e^{-\Phi}}{1+T}\sqrt{
1+T+\frac{\eta^{\mu\nu}\partial_{\mu}T
\partial_{\nu}T}{T}} \ 
\end{equation}
does not correctly captures the physics of the
tachyon condensation in the linear dilaton 
background (\ref{Kutbos}). 
In particular, we have shown that there  exists
the rolling tachyon solution $T\sim e^{\beta t}$ only
for large $T$, however the relation between
$\beta$ and $V_0$ is not the same as the relation
coming from the condition of the marginality of
the boundary interaction $T=e^{\beta t}$.
Moreover we have also shown that the behaviour of
 components of the stress energy tensor calculated from
(\ref{Kutdil}) differs from the results given
in \cite{Karczmarek:2003xm} in rather dramatic way. 
We meant that these  facts suggest that in order to find 
the tachyon effective action in the linear
dilaton background that would have 
the rolling tachyon as its exact solution 
we should allow more general insertion of $\Phi$.

The second leading point in our proposal is the
requirement  that for constant $\Phi$ our
action reduces to (\ref{Kutact}). 
For that reason we conjecture that
the factor in the square root of 
(\ref{Kutact}) should be generalised in 
case of the linear dilaton background as
\begin{equation}\label{bdil}
\bb=1+g_se^{-\Phi}\left(
\frac{T^2}{2}+\eta^{\mu\nu}\partial_{\mu}
T\partial_{\nu}T-T\eta^{\mu\nu}\partial_{\mu}T
\partial_{\nu}\Phi\right) \ .
\end{equation}
First of all we see 
that this term reduces to the standard
one given in (\ref{Kutact}) for $\Phi=\Phi_0$.
On the other hand for the half S-brane solution
\begin{equation}\label{halfS}
T=e^{\beta_{\mu}x^{\mu}} \ , 
\eta^{\mu\nu}\beta_{\mu}\beta_{\nu}    
-\eta^{\mu\nu}\beta_{\mu}V_{\nu}=\frac{1}{2}
\end{equation}
 (\ref{bdil}) is constant which
according to our observations given in 
 \cite{Kluson:2003sr} is an important fact for
the search of the exact solution with the
correct marginal condition. We have
also included  the exponential factor $e^{-\Phi}$
in (\ref{bdil}) from the following reason. 
If we  demand that
the action should have the tachyon profile
(\ref{halfS}) as its exact solution it is possible
that some  terms in the equation
of motion will be equal for
 $\sqrt{\bb}=1$ which, however need not
to be true for any $\bb=const\neq 1$.
 It would then be 
useful to have such an exact solution which
will lead to any constant value of $\bb$. In case
of the constant dilaton background such a solution
is full S-brane profile. Then the requirement that
the tachyon effective action should have this 
profile as its exact solution was used in the
determination of the tachyon effective  action in
\footnote{For former papers where such an approach
was used, see \cite{Lambert:2002hk,Lambert:2001fa}.}
\cite{Kutasov:2003er,Smedback:2003ur,Niarchos:2004rw}.
An analogue such a solution in case of the linear
dilaton background is 
\begin{equation}\label{Tpro}
T=T_+e^{\beta_{\mu}^+x^{\mu}}+T_-e^{\beta_{\mu}^-
x^{\mu}} \ ,
\beta_{\mu}^{\pm}\eta^{\mu\nu}\beta_{\nu}^{\pm}
-\eta^{\mu\nu}\beta_{\mu}^{\pm}V_{\nu}=\frac{1}{2}
 \ ,
\end{equation}
where now 
\begin{equation}
(\beta_0^{\pm})^2-\beta_0^{\pm}V_0=\frac{1}{2} 
\Rightarrow
\beta_0^{\pm}=\frac{V_0\pm\sqrt{V_0^2+2}}
{2} \ , \beta_i^+=V_i \ , \beta_i^-=0 \ .
\end{equation}
In this case we have
\begin{eqnarray}
\bb=1+g_se^{-\Phi}\left(\frac{1}{2}\left[
T_+^2e^{2\beta_{\mu}^+x^{\mu}}+2T_+T_-
e^{(\beta_{\mu}^++\beta_{\mu}^-)x^{\mu}}
+T_-^2e^{2\beta_{\mu}^-x^{\mu}}\right]+
\right.\nonumber \\
\left.+
T_+^2e^{2\beta_{\mu}^+x^{\mu}}
(\beta_{\mu}^{+}\eta^{\mu\nu}\beta_{\nu}^{+}
-\eta^{\mu\nu}\beta_{\mu}^{+}V_{\nu})+
T_-^2e^{2\beta_{\mu}^-x^{\mu}}
(\beta_{\mu}^{-}\eta^{\mu\nu}\beta_{\nu}^{-}
-\eta^{\mu\nu}\beta_{\mu}^{-}V_{\nu})+
\right. \nonumber \\
\left. +T_+T_-e^{(\beta_{\mu}^++\beta_{\mu}^-)
x^{\mu}}(2\beta_{\mu}^+\eta^{\mu\nu}
\beta_{\nu}^-
-(\beta_{\mu}^++\beta_{\mu}^-)\eta^{\mu\nu}V_{\nu})
\right)= \nonumber \\
=1+2g_sT_+T_-(1-\frac{V_{\mu}V^{\mu}}{2})
\nonumber \\
\end{eqnarray}
using
\begin{eqnarray}
(\beta_{\mu}^++\beta_{\mu}^-)x^{\mu}=
V_0x^0+V_ix^i=\Phi \ , \nonumber \\
\beta_{\mu}^+\eta^{\mu\nu}\beta_{\nu}^-=
-\beta_0^+\beta_0^-=\frac{1}{2} \ ,
\nonumber \\
(\beta_{\mu}^++\beta_{\mu}^-)\eta^{\mu\nu}V_{\nu}=
-V_0^2+V_i^2=V_{\mu}V^{\mu} \ . \nonumber \\
\end{eqnarray}

Now 
 we come to   proposal considering
tachyon effective action in the linear 
dilaton background. 
We  require that for
constant dilaton this Lagrangian reduces to the
(\ref{Kutact}) and also we would
like to have a connection with the  formulation
of the Lagrangian evaluated on
the tachyon marginal profile given in 
\cite{Kluson:2004pj}.
For that reason we  propose following form of the
supersymmetric tachyon effective Lagrangian
\begin{eqnarray}\label{acsuppro}
\mathcal{L}=\frac{1}{g_s}\sqrt{\bb}\int_0^{\infty}
ds e^{-s-\frac{1}{2}g_se^{-\Phi}T^2\frac{s^F}{G}}
\equiv \frac{1}{g_s}\sqrt{\bb}\int_0^{\infty}ds e^{-D(x,s)} \ ,
\nonumber \\
D(x,s)=s+x\frac{s^{F(V)}}{G(V)} \ , x\equiv
\frac{g_se^{-\Phi}T^2}{2} \ , \nonumber \\
\end{eqnarray}
where $F,G$ are unknown functions of 
$V_{\mu}$ that should obey following conditions:  
$\lim_{V\rightarrow 0} F(V)=1 \ ,
\lim_{V\rightarrow 0}G(V)=1$ that imply that
(\ref{acsuppro}) reduces to  (\ref{Kutact}).
Our goal is to determine these functions $F,G$ 
so that the general tachyon profile (\ref{Tpro})
 will be solution
of the equation of motion that arises 
from (\ref{acsuppro}) 
\begin{eqnarray}
-\sqrt{\bb}g_se^{-\Phi}T\int_0^{\infty} ds e^{-D}\frac{s^F}{G}
+\frac{g_se^{-\Phi}T}{2\sqrt{\bb}}\int_0^{\infty} ds e^{-D}-
\frac{g_se^{-\Phi}\eta^{\mu\nu}\partial_{\mu}T
\partial_{\nu}\Phi}{2\sqrt{\bb}}\int_0^{\infty}  ds e^{-D}
-\nonumber \\
-\partial_{\mu}\left[\frac{g_se^{-\Phi}
\eta^{\mu\nu}\partial_{\nu}T}{\sqrt{\bb}}
\int_0^{\infty} ds
e^{-D}\right]
+\partial_{\mu}\left[\frac{
g_se^{-\Phi}T\eta^{\mu\nu}\partial_{\nu}\Phi 
}{2\sqrt{\bb}}\int_0^{\infty} ds e^{-D}\right]=0
 \ .
\nonumber \\
\end{eqnarray}
Using the fact that $\sqrt{\bb}=const$ and also
\begin{equation}
g_se^{-\Phi}[\eta^{\mu\nu}\partial_{\mu}T\partial_{\nu}T
-T\eta^{\mu\nu}\partial_{\mu}T\partial_{\nu}\Phi]=\bb
-1-\frac{g_se^{-\Phi}T^2}{2} \ . 
\end{equation}
the equation of motion
(\ref{acsuppro})  after some calculation takes
the form
\begin{eqnarray}\label{eqf}
\int ds e^{-D}\frac{s^F}{G}\left[1+x\left(
1-\frac{V_{\mu}V^{\mu}}{2}\right))
\right]=
\int ds e^{-D}\left(1-\frac{V_{\mu}V^{\mu}}{2}\right) \ . \nonumber \\
\end{eqnarray}
The  last equation is obeyed if we  take   
\begin{equation}
F=1  \ ,
 \frac{1}{G}=(1-\frac{V_{\mu}V^{\mu}}{2}) 
\end{equation}
as can be seen from following integrals
\begin{eqnarray}
\int_0^{\infty}
 ds e^{-D}\frac{s^F}{G}\left[1+x\left(
1-\frac{V_{\mu}V^{\mu}}{2}\right)\right]=
\frac{(1-\frac{V_{\mu}V^{\mu}}{2})}{1+x(1-\frac{V_{\mu}V^{\mu}}{2})}
 \ , \nonumber \\
\left(1-\frac{V_{\mu}V^{\mu}}{2}\right)\int_0^{\infty} ds e^{-D}=
\frac{(1-\frac{V_{\mu}V^{\mu}}{2})}
{1+x(1-\frac{V_{\mu}V^{\mu}}{2})} \ . \nonumber \\
\end{eqnarray}
We see that the generalisation of 
(\ref{Kutact}) to the case of linear dilaton
background takes remarkable simple form
\begin{equation}\label{kudil2}
\mathcal{L}=\frac{1}
{g_s(1+\frac{g_se^{-\Phi}T^2}{2}(1-\frac{V_{\mu}V^{\mu}}{2}))}
\sqrt{\bb} \ . 
\end{equation}
Note that for  small $T$ (\ref{kudil2})
is equal to
\begin{eqnarray}\label{eflagst}
\mathcal{L}
=\frac{1}{g_s}-\frac{e^{-\Phi}T^2}{4}(1-V_{\mu}V^{\mu})+\frac{e^{-\Phi}
\eta^{\mu\nu}\partial_{\mu}T\partial_{\nu}T}{2}-\frac{e^{-\Phi}
T^2\partial_{\mu}\Phi\partial_{\nu}\Phi}{4}=
\nonumber \\
=\frac{1}{g_s}
+\frac{e^{-\Phi}}{2}\left[
\eta^{\mu\nu}\partial_{\mu}T\partial_{\nu}T
-\frac{1}{2}T^2\right] 
\nonumber \\
\end{eqnarray}
which is correct (up to constant term)
 tachyon effective Lagrangian around the point $T=0$. 
In the previous equation we have used  the integration 
by parts in the  D-brane
effective action that implies
\begin{eqnarray}
-\frac{1}{2}\int d^Dx
e^{-\Phi}T\eta^{\mu\nu}\partial_{\mu}T\partial_{\nu}\Phi=
-\frac{1}{4}\int d^Dx e^{-\Phi}T^2\eta^{\mu\nu}\partial_{\mu}\Phi
\partial_{\nu}\Phi \ . \nonumber \\
\end{eqnarray}
Tachyon effective action (\ref{kudil2})  also implies
one remarkable fact that could be useful for
further research. 
As is well known the norm of the dilaton
field  $V_{\mu}V^{\mu}$ is related to the
number of space-time dimensions as 
\cite{Polchinski:rq,Polchinski:rr}
\begin{equation}
V_{\mu}\eta^{\mu\nu}V_{\nu}=
\frac{10-D}{4} \ . 
\end{equation}
Then we immediately get that  the potential  
term in (\ref{kudil2})  vanishes
in two dimensions since
$1-\frac{V_{\mu}V^{\mu}}{2}=\frac{D-2}{8}$. We can interpret
this property as a consequence of the absence of   the
 perturbative instability
in two dimensional field theory where the tachyon is massless.

As a next step we will
 calculate the stress energy tensor  and the dilaton
source from (\ref{kudil2}). 
Since it is convenient to have the stress energy tensor symmetric
we rewrite the term containing gradient of the dilaton in
(\ref{kudil2}) as 
\begin{equation}
g_se^{-\Phi}\eta^{\mu\nu}\partial_{\mu}T\partial_{\nu}\Phi=
\frac{g_se^{-\Phi}\eta^{\mu\nu}}{2}
(\partial_{\mu}T\partial_{\nu}\Phi+\partial_{\mu}\Phi
\partial_{\nu}T) \ . 
\end{equation}
Before explicit calculation of the stress energy tensor
and dilaton charge we should discus one important
issue.  The effective tachyon action contains
the term $1-\frac{V_{\mu}\eta^{\mu\nu}V_{\nu}}{2}$
that for linear dilaton background 
can be written as  
\begin{equation}
1-\frac{\partial_{\mu}\Phi
\eta^{\mu\nu}\partial_{\nu}\Phi}{2} \ . 
\end{equation}
We mean that this is the right expression in
the action that we  should  used if we calculate
the  stress energy tensor and dilaton charge 
by variation of the action with respect to
$g^{\mu\nu}$ and $\Phi$.
 For
example, when we calculate the stress energy tensor
in flat space-time we replace the Minkowski metric
$\eta_{\mu\nu}$ with arbitrary $g_{\mu\nu}$, perform
the variation with respect to $g_{\mu\nu}$ and then
we again introduce the flat metric $\eta_{\mu\nu}$. 
With analogue with the previous example
we mean that it is appropriate to replace $V_{\mu}V^{\mu}$
with $g^{\mu\nu}
\partial_{\mu}\Phi\partial_{\nu}\Phi$, 
perform the variation with
respect to $\Phi$ and $g^{\mu\nu}$ and then insert
the values of metric and dilaton corresponding to the
flat space-time and linear dilaton background
\footnote{There could be potential subtlety with
this approach. Since we have
obtained the action (\ref{kudil2}) in fixed
linear dilaton background this off-shell
continuation could be ambiguous.}. As a result
of this calculation  we  get components of the stress
energy tensor for unstable D-brane in
linear dilaton background
\begin{eqnarray}
T_{\mu\nu}=-\eta_{\mu\nu}\mathcal{L}+2\frac{\delta
\mathcal{L}}{\delta g^{\mu\nu}}=\nonumber \\
-\frac{\eta_{\mu\nu}\sqrt{\bb}}{
g_s(1+\frac{g_se^{-\Phi}T^2}{2}(1-\frac{V_{\mu}V^{\mu}}{2}))}
+\frac{g_se^{-\Phi}\partial_{\mu}T\partial_{\nu}T-\frac{g_se^{-\Phi}T}{2}(
\partial_{\mu}TV_{\nu}+V_{\mu}
\partial_{\nu}T)}
{g_s(1+\frac{g_se^{-\Phi}T^2}{2}(1-\frac{V_{\mu}V^{\mu}}{2}))
\sqrt{\bb}}+ \nonumber \\
+\frac{g_se^{-\Phi}T^2V_{\mu}
V_{\nu}\sqrt{\bb}}{2g_s(1+g_se^{-\Phi}\frac{T^2}{2}
(1-\frac{V_{\mu}V^{\mu}}{2}))^2}
 \ . \nonumber \\
\end{eqnarray}
For the half S-brane solution with $T=e^{\beta x^0}$ we obtain
following components 
\begin{eqnarray}\label{halfT}
T_{00}=\frac{1+\frac{g_se^{-\Phi}T^2}{2}}{g_s(1+\frac{g_se^{-\Phi}T^2}{2}
(1-\frac{V_{\mu}V^{\mu}}{2}))}
+\frac{g_se^{-\Phi}T^2V_{0}
V_{0}}{2g_s(1+g_se^{-\Phi}\frac{T^2}{2}
(1-\frac{V_{\mu}V^{\mu}}{2}))^2}
 \ , \nonumber \\
T_{ij}=-\delta_{ij}\frac{1}{g_s(1+\frac{g_se^{-\Phi}T^2}{2}
(1-\frac{V_{\mu}V^{\mu}}{2}))} 
+\frac{g_se^{-\Phi}T^2V_{i}
V_{j}}{2g_s(1+g_se^{-\Phi}\frac{T^2}{2}
(1-\frac{V_{\mu}V^{\mu}}{2}))^2}
 \ , \nonumber \\
T_{0i}=T_{i0}=-\frac{g_se^{-\Phi}\beta V_iT^2}{2g_s(
 1+\frac{g_se^{-\Phi}T^2}{2}
(1-\frac{V_{\mu}V^{\mu}}{2}))}
+\frac{g_se^{-\Phi}T^2V_{0}
V_{i}}{2g_s(1+g_se^{-\Phi}\frac{T^2}{2}
(1-\frac{V_{\mu}V^{\mu}}{2}))^2}
 \ . \nonumber \\
\end{eqnarray}
In the same way we get the dilaton source
\begin{eqnarray}
J_{\Phi}=-\frac{\delta \mathcal{L}}{\delta \Phi}=
-\frac{g_se^{-\Phi}T^2(1-\frac{V_{\mu}V^{\mu}}{2})\sqrt{\bb}}
{2g_s( 1+\frac{g_se^{-\Phi}T^2}{2}
(1-\frac{V_{\mu}V^{\mu}}{2}))^2}
+\frac{g_se^{-\Phi}\left(\frac{T^2}{2}+\eta^{\mu\nu}
\partial_{\mu}T\partial_{\nu}T-\eta^{\mu\nu}\partial_{\mu}T
V_{\nu}\right)}{2g_s\sqrt{\bb}
( 1+\frac{g_se^{-\Phi}T^2}{2}
(1-\frac{V_{\mu}V^{\mu}}{2}))}-\nonumber \\
-\partial_{\mu}\left[\frac{g_se^{-\Phi}T\eta^{\mu\nu}
\partial_{\nu}T}{2g_s(1+\frac{g_se^{-\Phi}T^2}{2}
(1-\frac{V_{\mu}V^{\mu}}{2}))\sqrt{\bb}}\right]+ 
\partial_{\mu}
\left[\frac{\eta^{\mu\nu}V_{\nu} 
g_se^{-\Phi}T^2\sqrt{\bb}}{2g_s(1+g_se^{-\Phi}\frac{T^2}{2}
(1-\frac{V_{\mu}V^{\nu}}{2}))^2}\right]
 \nonumber \\
\end{eqnarray}
that for the tachyon profile $T=e^{\beta x^0}$
is equal to
\begin{eqnarray}
J_{\Phi}
=-\frac{g_se^{-\Phi}T^2(-\frac{V_{\mu}V^{\mu}}{2}-\beta V_0)}
{2g_s( 1+\frac{g_se^{-\Phi}T^2}{2}
(1-\frac{V_{\mu}V^{\mu}}{2}))^2} 
+\frac{(-\beta V_0-\frac{V_{\mu}V^{\mu}}{2})
g_se^{-\Phi}T^2(1-\frac{g_se^{-\Phi}T^2}{2}(
1-\frac{V_{\mu}V^{\mu}}{2}))}
{g_s(1+g_se^{-\Phi}\frac{T^2}{2}
(1-\frac{V_{\mu}V^{\nu}}{2}))^3} \ . \nonumber \\
\end{eqnarray}
One can  easily confirm that $T_{\mu\nu}$ and
$J_{\Phi}$ are related through the conservation
law
\begin{equation}
\partial^{\mu}T_{\mu\nu}=
V_{\nu}J_{\Phi} \  .
\end{equation}
From (\ref{halfT}) we determine 
the asymptotic behaviour 
of the components  of the stress energy tensor at
far future and far past. Using the fact that
for the half S-brane $e^{-\Phi}T^2=
e^{-V_ix^i+t\sqrt{V_0^2+2}}$ we 
obtain the asymptotic behaviour of $T_{\mu\nu}$
at far future
\begin{eqnarray}
T_{00}
\sim \frac{1}{g_s\left(1-\frac{V_{\mu}V^{\mu}}{2}\right)} \ ,
T_{ij}
\sim 0 \ ,
T_{0i}
\sim -\frac{\beta V_i}{g_s(1-\frac{V_{\mu}V^{\mu}}{2})}
\nonumber \\
\end{eqnarray}
together with
\begin{equation}
J_{\Phi}\rightarrow 0 \ .
\end{equation}
We see that at far future the energy density is independent
on time while $T_{ij}$ scales to zero. It would be nice to
interpret this result as a fact, that the the tachyon dust
is compression of gas of massive closed string whose energy
is unaffected in a way by change of the dilaton. This behaviour
is in agreement with the CFT calculation given in
\cite{Karczmarek:2003xm}
On the other hand CFT analysis implies
that off-diagonal components of the stress
energy tensor are zero while we have got
nonzero components  $T_{0i}$ whose meaning
is at present not completely 
clear to us.  However the fact that they
are proportional to $-V_i$ suggests that tachyon dust
slides towards weak coupling which would be in agreement
with the interpretation given in \cite{Karczmarek:2003xm}.
To see more clearly this correspondence 
we should  study the fluctuations around different tachyon rolling
solution. We hope to return to these problems in future. 
However much more serious problem arises in the opposite
limit $t\rightarrow -\infty$. In this case we again find
\begin{eqnarray}
T_{00}\sim \frac{1}{g_s}+\frac{g_se^{-\Phi} T^2}{g_s4}V_{\mu}V^{\mu}
+\frac{g_se^{-\Phi}T^2V_0^2}{g_s2}\rightarrow \frac{1}{g_s} 
 \ , \nonumber \\
T_{ij}\sim  -\delta_{ij}\left(1-\frac{g_se^{-\Phi} T^2}{g_s2}
\left(1-\frac{V_{\mu}V^{\mu}}{2}\right)
\right)+\frac{g_se^{-\Phi}T^2V_iV_j}{g_s2}
\rightarrow -\delta_{ij}\frac{1}{g_s} 
\ , \nonumber \\
T_{0i}=-\frac{g_se^{-\Phi}T^2V_i}{g_s2}
\left(\beta-V_0\right)\rightarrow 0 \ , \nonumber \\
J_{\Phi}\sim \frac{g_se^{-\Phi}T^2}{g_s} \rightarrow 0 \ .  \nonumber \\
\end{eqnarray}
Even if  the components of the stress energy tensors
scale as $\frac{1}{g_s}$ this result is not in agreement with
the calculations given in   \cite{Karczmarek:2003xm}
where it was shown that components of the stress
energy tensors scale at far past as $\sim e^{-\Phi}$.

 We mean that this could be indication
that our proposed effective action, even if it has
the tachyon rolling solution for every values of
$T$, should be modified at small $T$ where 
we could expect that terms with higher derivatives
could be important. For example, let us presume
that some additional term in the action contains
following combination $\mathbf{C}=1+\frac{T^2}{2}
+T\partial_{\mu}\eta^{\mu\nu}\partial_{\nu}T
-T\eta^{\mu\nu}\partial_{\mu}T\partial_{\nu}\Phi$. 
Clearly this combination is equal to one for any 
marginal tachyon profile and hence the term in
the  tachyon effective Lagrangian that contains such a
combination should come with the standard factor $e^{-\Phi}$
and consequently could give significant contribution to
the tachyon dynamics in weak coupling. On
the other hand when we include terms containing
the second derivatives to the action there is 
no reason why we should not include terms containing
derivative of higher orders. For that reason
we restrict ourselves in our proposal to the
action containing first derivatives only and
leave the more general case for future work.
\section{Proposal for tachyon effective
action in bosonic string theory in
linear dilaton background}\label{third}
In this section we will propose the 
generalised form of the tachyon effective
action in bosonic string theory in
linear dilaton background. Recall that in
the flat space-time the tachyon effective
action has form given in (\ref{Kutbos}).
Following the results given in case of the
supersymmetric action we should include 
some powers of the
factor $g_se^{-\Phi}$ in front of the tachyon $T$.
Since in the effective action (\ref{Kutbos}) tachyon
components scales as $T$ we propose that we
should include everywhere the factor 
$g_s^{1/2}e^{-\Phi/2}$. In other words our proposal
for the tachyon effective action in bosonic
string theory has the form
\begin{eqnarray}\label{bospro}
\mathcal{L}=\frac{1}{g_s}\sqrt{\bb}\int_0^{\infty}ds
e^{-s-g_s^{1/2}e^{-\Phi/2}T \frac{s^{F}}{G}}\equiv
\sqrt{\bb}\int_0^{\infty}ds e^{-D} \ , \nonumber \\
\bb=1+g_s^{1/2}e^{-\Phi/2}\left[
\frac{\eta^{\mu\nu}\partial_{\mu}T\partial_{\nu}T}{T}
-\eta^{\mu\nu}\partial_{\nu}T\partial_{\nu}\Phi\right] \ ,
 \nonumber \\ 
 D=s+x\frac{s^F}{G} \ , x\equiv g_s^{1/2}e^{-\Phi/2}T \ . 
\nonumber \\
\end{eqnarray}
with the property that for $V\rightarrow 0$ we
have $F \ ,G\rightarrow 1$. The equation of motion
that arises from (\ref{bospro}) is
\begin{eqnarray}
-g_s^{1/2}e^{-\Phi/2}\sqrt{\bb}\int_0^{\infty}
 ds e^{-D}\frac{s^F}{G}  
+\frac{g_s^{1/2}e^{-\Phi/2}}{2\sqrt{\bb}}\int_0^{\infty} ds e^{-D}
-g_s^{1/2}e^{-\Phi/2}
\frac{\eta^{\mu\nu}
\partial_{\mu}T\partial_{\nu}T}{2T^2\sqrt{\bb}}
\int_0^{\infty} ds e^{-D}
-\nonumber \\
- \partial_{\mu}\left[
\frac{g_s^{1/2}e^{-\Phi/2}
\eta^{\mu\nu}\partial_{\nu}T}{T\sqrt{\bb}}
\int_0^{\infty} ds e^{-D}\right]
+\partial_{\mu}\left[
\frac{g_s^{1/2}e^{-\Phi/2}
\eta^{\mu\nu}\partial_{\nu}\Phi}{2\sqrt{\bb}}
\int_0^{\infty} ds e^{-D}\right]
=0  \nonumber \\
\end{eqnarray}
that for the tachyon profile $T=e^{\beta_{\mu}x^{\mu}} \ ,
\bb=1$ is equal to
\begin{equation}
g_s^{1/2}e^{-\Phi/2}\int ds e^{-D}\frac{s^F}{G}
\left[1+x(1-\frac{V^2}{4})\right]=
g_s^{1/2}e^{-\Phi/2}\int ds e^{-D}
\left(1-\frac{V^2}{4}\right) \ . 
\end{equation}
The last formulation suggests that $F,G$ should
be equal to
\begin{equation}
F=1 \ , \frac{1}{G}=
(1-\frac{V_{\mu}V^{\mu}}{4}) \ 
\end{equation}
as follows from the same arguments as
were  given in the previous section.
Since in the linear dilaton background 
we have $\partial_{\mu}\Phi=V_{\mu}$ we can
also write the tachyon effective action as
\begin{equation}\label{actbose}
\mathcal{L}=\frac{\sqrt{\bb}}{g_s}
\int_0^{\infty}ds e^{-s(1+\frac{x}{G})}=
\frac{1}{g_s\left(1+g_s^{1/2}e^{-\Phi/2}T(1-\frac{\eta^{\mu
\nu}\partial_{\mu}\Phi
\partial_{\nu}\Phi}{4})\right)} \ . 
\sqrt{\bb}
\end{equation}
Now we will calculate  the stress 
energy tensor from (\ref{actbose}). This calculation
can be performed completely in the same way
as in previous section with the result
\begin{eqnarray}
T_{\mu\nu}=-\eta_{\mu\nu}\mathcal{L}+2\frac{\delta
\mathcal{L}}{\delta g^{\mu\nu}}=\nonumber \\
=-\eta_{\mu\nu}
\frac{1}{g_s(1+e^{-\Phi/2}T(1-\frac{V_{\mu}V^{\mu}}{4}))}
\sqrt{\bb}+\nonumber \\
+\frac{\frac{g_s^{1/2}e^{-\Phi/2}
\partial_{\mu}T\partial_{\nu}T}{T}-\frac{g_s^{1/2}e^{-\Phi/2}}{2}
(\partial_{\mu}T\partial_{\nu}\Phi+\partial_{\mu}\Phi
\partial_{\nu}T)}{g_s(1+g_s^{1/2}e^{-\Phi/2}T(1-\frac{V_{\mu}V^{\mu}}{4}))
\sqrt{\bb}}+
 \nonumber \\
+\frac{ g_se^{-\Phi/2}TV_{\mu}V_{\nu}\sqrt{\bb}}
{2g_s(1+g_s^{1/2}e^{-\Phi/2}T(1-\frac{V_{\mu}V^{\mu}}{4}))^2}
\nonumber \\
\end{eqnarray}
that for the 
 rolling tachyon solution $T=e^{\beta x^0}$  gives
\begin{eqnarray}
T_{00}=\frac{1+g_s^{1/2}e^{-\Phi/2}T}
{g_s(1+g_s^{1/2}e^{-\Phi/2}T(1-\frac{V_{\mu}
V^{\mu}}{4}))} 
+\frac{ g_s^{1/2}e^{-\Phi/2}TV_0^2}
{2g_s(1+g_s^{1/2}e^{-\Phi/2}T(1-\frac{V_{\mu}V^{\mu}}{4}))^2} \ , 
\nonumber \\
T_{ij}=-\delta_{ij}\frac{1}{g_s(1+g_s^{1/2}e^{-\Phi/2}T(1-\frac{V_{\mu}
V^{\mu}}{4})}
+\frac{ g_s^{1/2}e^{-\Phi/2}TV_{i}V_{j}}
{2g_s(1+g_s^{1/2}e^{-\Phi/2}T(1-\frac{V_{\mu}V^{\mu}}{4}))^2}
\nonumber \\
 \ , 
\nonumber \\
T_{0i}=T_{i0}=-\frac{\beta_0V_ig_s^{1/2}e^{-\Phi/2}T}
{2g_s(1+g_s^{1/2}e^{-\Phi/2}T(1-\frac{V_{\mu}
V^{\mu}}{4}))}
+\frac{ g_s^{1/2}e^{-\Phi/2}TV_{0}V_{i}}
{2g_s(1+g^{1/2}_se^{-\Phi/2}T(1-\frac{V_{\mu}V^{\mu}}{4}))^2}
 \ . 
\nonumber \\
\end{eqnarray}
For $t\rightarrow \infty$ we have
\begin{equation}
T_{00}\rightarrow \frac{1}{g_s(1-\frac{V_{\mu}V^{\mu}}{4})} \ ,
T_{ij}\rightarrow 0 \ ,T_{0i}\rightarrow
-\frac{\beta_0V_i}{g_s(1-\frac{V_{\mu}V^{\mu}}{4})} \ ,
\end{equation}
while for $t\rightarrow -\infty$ we again obtain the same behaviour
as in supersymmetric case given in previous section
\begin{equation}
T_{\mu\nu}\sim -\eta_{\mu\nu}\frac{1}{g_s} \ .
\end{equation}
After some straightforward calculation we can 
show that the stress energy tensor obeys the conserved
law 
\begin{equation}\label{conserv}
\partial^{\mu}T_{\mu\nu}=V_{\nu}J_{\Phi} \ ,
\end{equation}
where 
 the dilaton source $J_{\Phi}=-\frac{\mathcal{L}}{\delta 
\Phi}$  is equal to 
\begin{eqnarray}
J_{\Phi}=\frac{e^{-\Phi/2}T(1-\frac{V_{\mu}V^{\mu}}{4})}
{2(1+e^{-\Phi/2}T(1-\frac{V_{\mu}
V^{\mu}}{4}))^2}\sqrt{\bb}-
\frac{e^{-\Phi/2}(T+\frac{\eta^{\mu\nu}\partial_{\mu}T
\partial_{\nu}T}{T}-\eta^{\mu\nu}\partial_{\mu}T\partial_{\nu}\Phi)
}{4(1+e^{-\Phi/2}T(1-\frac{V_{\mu}
V^{\mu}}{4}))\sqrt{\bb}}+\nonumber \\
+\partial_{\mu}\left[\frac{e^{-\Phi/2}\eta^{\mu\nu}\partial_{\nu}T}
{2(1+e^{-\Phi/2}T(1-\frac{V_{\mu}
V^{\mu}}{4}))\sqrt{\bb}}\right]
-\partial_{\mu}\left[\frac{Te^{-\Phi/2}\eta^{\mu\nu}
\partial_{\nu}\Phi\sqrt{\bb}}
{2(1+e^{-\Phi/2}T(1-\frac{V_{\mu}
V^{\mu}}{4}))^2}\right] \ .
\nonumber \\
\end{eqnarray}
In this section we have performed
 the generalisation
of the tachyon effective action (\ref{Kutbos})
to the case of linear dilaton field. As in
supersymmetric case we have got  the tachyon
effective action that has remarkable simple form
and that has the rolling tachyon in the linear
background as its exact solution. On the
other hand we have
also seen that there are problems with the
asymptotic behaviour of the stress energy
tensors whose origin  is probably the
same as in the supersymmetric case. 
\section{Conclusion}\label{fourth}
In this paper we  proposed 
the tachyon effective actions for unstable
space-time filling D-branes in
superstring and bosonic theories 
in the linear dilaton background.
Our proposal was based on requirements
 that for constant
$\Phi$ these tachyon effective actions reduce
to the tachyon effective actions 
(\ref{Kutact}) and (\ref{Kutbos}). The second condition
was that in the case of unstable D-brane 
in superstring theory the rolling tachyon profiles
 known as half S-brane and full S-brane are solutions 
of the equation of motion that arises from it. 
Using these conditions we were able 
to obtain the tachyon effective action  for unstable
space-time filling D-brane in linear dilaton background.
Then we apply the similar procedure to the
case of tachyon effective action in bosonic
theory and we have found the action that
has half S-brane in the linear dilaton background
as its exact solution
\footnote{We have restricted ourselves to the
half S-brane tachyon profile since the action
(\ref{Kutbos}) has not solution corresponding to
the full S-brane, for more general approach to
the tachyon effective actions in bosonic theory,
see \cite{Smedback:2003ur}.}.
 We  have also calculated  the stress
energy tensors from these actions and
then we have studied their  behaviour for
half S-brane tachyon profile with emphasise
to their asymptotic form at far past and
future. 
 In the far future we have found that
the energy remains constant and the pressure goes 
to zero together with the observation that the
tachyon dust seems to drift to the weak coupling region.
We mean that these results are in rough
coincidence  with the calculations performed
in  \cite{Karczmarek:2003xm} even if the stress
energy tensor calculated here is different from
the stress energy tensor given in \cite{Karczmarek:2003xm}.
This fact is not surprising since as was shown
recently in \cite{Fotopoulos:2003yt,Kluson:2004pj}
tachyon effective actions (\ref{Kutact}) and
(\ref{Kutbos}) are not directly determined from
the string partition function.
 On the other hand we have seen
that in the far past the behaviour of the
stress energy tensor is completely different
from results presented in \cite{Karczmarek:2003xm}
that show that in the asymptotic past the
components of the stress energy tensor scale
as $\frac{1}{g_s(t)}$. On the other hand
we have found that in this limit the diagonal 
components of the
stress energy tensor approach  constant 
values and the off-diagonal
ones vanish.  It seems to us 
that this result is uncomfortable and
suggests that something important 
is missing in our analysis. 
We mean that in order to correctly describe
the tachyon dynamics near the point $T=0$ we 
should include into 
the effective Lagrangians terms that contain
higher order derivatives of tachyon and 
which  should give significant
contributions around the point $T=0$.
 In fact there is no reason why
we should restrict ourselves to the action
with the first derivatives only since as 
one can see all higher derivatives scale
as the first order one in case of rolling
tachyon profile. We must also stress
that the proposed form of the tachyon effective
action was not determined 
directly from the string theory analysis that should
be based on the calculation using the powerful machinery of 
background independent string field theory
\cite{Witten:1992qy}
 or alternatively 
using the sigma model approach \cite{Tseytlin:2000mt}.

In spite of these comments we still hope
 that our proposed actions 
 could be useful
for the study of some aspects of
 D-branes in two dimensions
 \cite{Fateev:2000ik,
Teschner:2000md,Zamolodchikov:2001ah,
Schomerus:2003vv,Teschner:2003qk}. For
that reason we  mean that it deserves 
further study, at least as a toy
model for the description of the
tachyon condensation on D-brane in nontrivial
dilaton background. 
So that let us mention  some possibilities
for further research. We mean  
that it would be interesting to 
study how the tachyon effective
action given above could be applied 
in the tachyon cosmology (For
recent interesting papers discussing
this subject 
with detailed list of references, see
\cite{Gibbons:2003gb,Sen:2003mv}.) It would be also 
nice to see whether our
proposal could be useful in the study 
of D-branes decay in two dimensions
\cite{McGreevy:2003kb,Klebanov:2003km,McGreevy:2003ep,
Sen:2003iv,Gutperle:2003ij,Douglas:2003up,Takayanagi:2003sm,
Klebanov:2003wg,DeWolfe:2003qf,Karczmarek:2003pv,Giveon:2003wn,
Kapustin:2003hi} even if  there are
limitations in application of our action 
since it does not contain the
coupling between the closed string tachyon and D-brane.
 We also believe
that our action could be useful in 
recent analysis of  two dimensional
Type OA and Type 0B  theories
\cite{Gross:2003zz,Danielsson:2003yi,Gukov:2003yp,
Strominger:2003tm,Yin:2003iv,Ho:2004qp,Takayanagi:2004jz,
Danielsson:2004xf}. We hope to return
to these problems in near future. 
\\
\\
{\bf Acknowledgement}
This work was supported by the
Czech Ministry of Education under Contract No.
14310006.
\\

\end{document}